\documentstyle [12pt] {article}
\newcommand{\cs}[3]{{{#3} \brace {#1 #2}}}
\begin{document}
\title {The Bazanski Approach in Brane Worlds: \\ A Brief Introduction}
\maketitle
\begin{center}
{\bf {M.E.Kahil{\footnote{Mathematics Department, Modern Sciences and Arts University, Giza, EGYPT\\
e.mail: kahil@aucegypt.edu}}} }
\end{center}
\begin{center}
Paths of test particles, rotating  and charged objects in brane-worlds  using a modified Bazanski Lagrangian are derived. We also discuss the transition to their corresponding equations in four dimensions. We then make a  comparison between the given equations in brane-worlds (BW) and their analog in space-time-matter (STM) theory.
\end{center}

{\bf{1.The Bazanski Approach in 5D}} \\
Motion of test particles in higher dimensions is obtained by using the usual Bazanski Lagrangian [1] which has the advantage that we obtain path and path deviation equations from the same Lagragian: 
\begin{equation}
L = g_{_{AB}}{U^{A}} {\frac{D \Psi^{B}}{DS}}
\end{equation}
where $A=1,2,3,4,5$.
By taking the variation with respect to the deviation vector $\Psi^{C}$ and the tangent vector $U^{C}$, we obtain the well known geodesic and geodesic deviation equations respectively:
\begin{equation}
\frac{dU^{C}}{dS}+ \cs{A}{B}{C} {U ^{A}}{U^{B}}=0
\end{equation} 
\begin{equation}
\frac{D^2 \Psi^{C}}{DS^2}= R^{C}_{ABD}U^{A}U^{B}\Psi^{D}
\end{equation}
Recently, the Bazanski Lagrangian has been modified in order to describe motion of charged particles and rotating objects in 5-dimensions whether they be compact or noncompact spaces [2]:  \\
\underline{In Compact Spaces}\\
The process to unify electromagnetism (gauge fields) and gravity
depends on extra component(s) of the metric. Using the cylinder condition, a charged particle whose behavior is described by the Lorentz equation in 4D behaves as
a test particle moving on a geodesic in 5D. At the same time, its deviation equation becomes like the well known geodesic deviation equation [2]. This result is obtained from applying the usual Bazanski method in 5D.\\
\underline{In Non-Compact Spaces}\\
the path equation has two main defects: \\
{{(i)} it is not gauge invariant} , \\{{(ii)} the additional extra force from an extra dimension is parallel to the four vector velocity}. \\
Some authors [4] and [5] have introduced different types of transformations in order overcome the above mentioned problems. These are expressed like the usual  geodesic equation (2). Applying the Bazanski approach we can obtain equation (2) and  its corresponding geodesic deviation equation, satisfying the Campbell-Magaard theorem [3]. Thus (2) becomes :
\begin{equation}
\frac{D^{2}\Psi^{A}}{DS^{2}}=0 .
\end{equation}

{\bf{2.The Bazanski Approach in Brane World Models}} \\
In the Brane world scenario our universe can be described in terms of 4+N dimensions with   $N \geq 1$  and the 4D space-time part of it  is embedded in 4+N. manifold [6]. Accordingly, the bulk geodesic motion is observed by a four dimensional 
observer to reproduce the physics of 4D space-time [7]. Consequently, the importance of the equation of motion for a test particle in the bulk space-time of brane worlds is to describe the apparant motion in 4D space-time.  Applying the Bazanski approach, we can obtain the  motion for a test particle  on a brane using the followinng Lagrangian :  
 \begin{equation}
L = g_{\mu \nu}(x^{\rho},y) U^{\mu} \frac{D \Psi ^{\nu}}{Ds} +f_{\mu}\Psi^{\mu},
\end{equation}
where $g_{\mu \nu}(x^{\rho},y)$ is the induced metric and$f_{\mu} =  \frac{1}{2} U^{\rho}U^{\sigma} \frac{\partial g_{\rho \sigma}}{\partial y}\frac{dy}{ds} U_{\mu}$ describes a parallel force due to the effect of non-compactified extra dimension to give [8]:
\begin{equation}
\frac{dU^{\mu}}{ds}+ \cs{\alpha}{\beta}{\mu} {U ^{\alpha}}{U^{\beta}} =  (\frac{1}{2}U^{\mu}U^{\sigma}-g^{\mu \sigma}) \frac{\partial g_{\rho \sigma}}{\partial y}\frac{dy}{ds} U^{\rho} .
\end{equation}
As in Brane world models, one can express $\frac{1}{2} \frac{\partial g_{\rho \sigma}}{\partial y}$ in terms of the extrinsic curvature $\Omega_{\rho \sigma}$ i.e.   
$ 
\Omega_{\alpha \beta} = \frac{1}{2} \frac{\partial g_{\rho \sigma}}{\partial y} 
$ [9].
Thus, the path equation for a test particle in brane world models becomes:
\begin{equation}
\frac{dU^{\mu}}{ds}+ \cs{\alpha}{\beta}{\mu} {U ^{\alpha}}{U^{\beta}} =2(\frac{1}{2}U^{\mu}U^{\sigma}-g^{\mu \sigma}) \Omega_{\rho \sigma} \frac{dy}{ds} U^{\rho}. 
\end{equation}  
Also,for a rotating object the Papapetrou equation [10] in 4 dimensions  becomes  :
\begin{equation}
\frac{dU^{\mu}}{ds}+ \cs{\alpha}{\beta}{\mu} {U ^{\alpha}}{U^{\beta}}= \frac{1}{m} \Omega^{\mu}  _{[ \gamma} \Omega_{\delta ] \beta} 
S^{\beta \gamma} U^{\delta} + 2(\frac{1}{2}U^{\mu}U^{\sigma}-g^{\mu \sigma})\Omega_{\rho \sigma} \frac{dy}{ds}  U^{\rho}.
\end{equation}

The above equation is derived from the following Lagrangian
\begin{equation}
L= g_{\mu \nu}(x^{\rho},y) U^{\mu} \frac{D \Psi^{\nu}}{Ds} + (\frac{1}{2m} R_{\mu \nu \rho \sigma} S^{\rho \sigma} U^{\nu} + \frac{1}{2} \frac{\partial g_{\rho \sigma }}{\partial y} U^{\rho}U^{\sigma} U_{\mu} \frac{dy}{ds} ) \Psi ^{\mu}, 
\end{equation}
with an additional factor related to Guass-Codassi equation (cf.[11]) and taking into consideration the Campbell-Magaard theorem, the four dimensional curvature becomes 
 i.e. 
 $
 R_{\alpha \beta \gamma \delta} = 2 \Omega_{\alpha[ \gamma} \Omega_{\delta]\beta}.  
 $ 
Similarly, the ususal equation of motion for a charged object [12] can be described in the presence of brane world models: 
 \begin{equation}
\frac{dU^{\mu}}{ds}+ \cs{\alpha}{\beta}{\mu} {U ^{\alpha}}{U^{\beta}}= \frac{q}{m} F^{\mu}_{. \beta} U^{\beta} + 2(\frac{1}{2}U^{\mu}U^{\sigma}-g^{\mu \sigma})\Omega_{\rho \sigma} \frac{dy}{ds}U^{\rho}, 
\end{equation}  
which is derived from the following Lagrangian: 
\begin{equation}
L= g_{\mu \nu}(x^{\rho},y) U^{\mu} \frac{D \Psi^{\nu}}{Ds} + (\frac{q}{m} F_{\mu \nu} U^{\nu} + \frac{1}{2} \frac{\partial g_{\rho \sigma }}{\partial y} U^{\rho}U^{\sigma} U_{\mu} \frac{dy}{ds} ) \Psi ^{\mu} 
\end{equation}

{\bf{3.Discussion and Concluding Remarks}} \\
In Brane world models, it can be easily found that matter in 4D is regarded as the effect of curvature of the extra dimension in a 5D bulk.
While in STM, the bulk is obtained due to the solution of 5D Einstein equations in vacuum [13]. This may show the equivalence between Brane world models and Space-time-matter theory as the first is embedding physics of 4D in order to describe the geometry of the bulk in 5D. While, in STM the process is based on projecting the geometry of a 5D flat curvature onto a 4D space to unify matter with geometry. From this perspective, it is worth mentioning that  equations of motions of a test paricle defined in STM (2), after pojecting the 5D equations onto 4D are equivalent to using their counter-part in brane worlds (6).  Each of them can be derived from a different Lagrangian using the Bazanski approach.   

 Also, in this work we have developed the equations of motion for  rotating and for charged objects while taking into account the effect of the extrinsic curvature on these sets of equations. Then equations (8) and (10)  reduce to their ususal rotating (Papapetrou)[10] and charged (Lorentz) equations (cf.[12]) respectively when the effect of the extra dimension is dropped.

Finally, it remains an important point should be discussed in future work:\\  
Is the Riemannian Geometry in higher dimensions (compact/non-compact) sufficient for  defining the physics of our cosmos?
This apprach might allow brane world models to be described using some non-symmetric geometries admitting non-vanishing curvature and torsion simulatenously. Perhaps these could lead a unification of all fields within the context of Brane world models.

{\bf{Acknowlgements}}\\
The author would like to thank  Professors M.I. Wanas, M. Abdel-Megied , G. De Young, V. Gurzadyan, K. Buchner,  T. Harko  and M. Roberts for their encourgement and comments. And a special word of thanks should be addressed to Professor R. Ruffini and the ICTP for their support and hospitality during the conference.  

{\bf{References}} \\
{[1]} Bazanski, S.L. (1989) J. Math. Phys., {\bf{30}}, 1018. \\
{[2]} Kahil, M.E. (2006), J. Math. Physics {\bf{47}},052501. \\
{[3]} Dahia, E., Monte,M. and Romaro,C. (2003),
Mod.Phys.Lett.{\bf{A18}},1773;

gr-qc/0303044 \\
{[4]} Ponce de Leon, J. (2002) Grav. Cosmology, {\bf{8}}, 272; gr-qc/ 0104008 \\
{[5]} Seahra, S.,(2002) Phys Rev. D. {\bf{65}}, 124004 gr-qc/0204032\\
{[6]} Liu, H. and Mashhoon, B. (2000), Phys. Lett. A, {\bf{272}},26 ;
 gr-qc/0005079 \\
{[7]} Youm, D. (2000) Phys.Rev.D{\bf{62}}, 084002; hep-th/0004144 \\ 
{[8]}Ponce de Leon, J. (2001) Phys Lett B, {\bf{523}} ;gr-qc/0110063 \\
{[9]}Dick, R.  Class. Quant. Grav.(2001), {\bf{18}}, R1. \\ 
{[10]} Papapetrou, A.(1951), Proc. Roy.Soc. Lond A{\bf{209}},248. \\
{[11]} Maartens, R. (2004) Living Rev.Rel.{\bf{7}}; gr-qc/0312059\\   
{[12]}Sen, D.K. , {\it{Fields/Particles}}, The Ryerson Press Toronto (1968). \\
{[13]} Ponce de Leon, J. (2001) Mod. phys. Lett. A {\bf{16}}, 2291
 ;gr-qc/0111011 \\

\end{document}